# Earthquake Response Analysis of Yielding Structures Coupled with Rocking Walls

M. Aghagholizadeh[1], N. Makris
*Southern Methodist University*

## ABSTRACT

This paper investigates the inelastic response of a yielding structure coupled with a rocking wall which can be vertically restrained. The paper first derives the nonlinear equations of motion of a yielding oscillator coupled with a vertically restrained rocking wall and the dependability of the one-degree of freedom idealization is validated against the nonlinear time-history response analysis of a well-known 9-story moment-resisting steel frame that is coupled with a stepping rocking wall. While, the coupling of weak building frames with rocking walls is an efficient strategy that controls inelastic deformations by enforcing a uniform interstory-drift distribution, therefore, avoiding mid-story failures, the paper shows that even for medium-rise buildings the effect of vertical tendons on the inelastic structural response is marginal, with the exception of increasing the vertical reactions at the pivoting points of the rocking wall. Accordingly, the paper, concludes that for medium- to high-rise buildings vertical tendons in rocking walls are not beneficial.

## INTRODUCTION

In an effort to eliminate the appreciable seismic damage in moment-resisting frames that occasionally resulted to a weak-story failure, the concept of a rigid core system gained appreciable ground (Paulay 1969, Fintel 1975, Emori et al. 1978, Bertero 1980, Aktan et al. 1984). When the core walls in tall buildings are fixed-based, the ductility capacity of the base of the core wall may be limited given the significant axial loads; while, the ductility demands are appreciable under long-duration pulse motions (Paulay 1986, Zhang et al. 2000). Furthermore, the base of the core wall may suffer from cyclic degradation under prolonged shaking which usually results to permanent inelastic deformations. Such inelastic response may result to permanent drifts and lead to large repair costs; therefore, the entire design becomes unsustainable.

During the last three decades, there has been a growing effort to direct the attention of engineers to the unique advantages associated with allowing major vertical structural elements (piers in bridges or shear wall in buildings) to uplift in an effort to intentionally mobilize a lower "failure" mechanism. In this way failures associated with cyclic degradation are essentially avoided; while, permanent displacements remain small due to the inherent recentering tendency of the rocking mechanism. For instance, as early as the PRESS Program (Priestley 1991, 1996), the jointed shear wall system was allowed to lift-off and rock (Nakaki et al. 1999, Priestley et al. 1999). About the same time Kurama et al. (1999, 2002) examined the lateral load behavior of unboded segmented post-tensioned precast walls; while, Mander and Cheng (1997) introduced the damage avoidance design (DAD) in which the free-standing piers of a bridge frame are only vertically restrained through their center line and are allowed to rock atop the pile-cap and bellow the pier-cap beam without inducing any damage.

---

[1] Corresponding Author: M. Aghagholizadeh, *Southern Methodist University, mehrdadag@smu.edu*

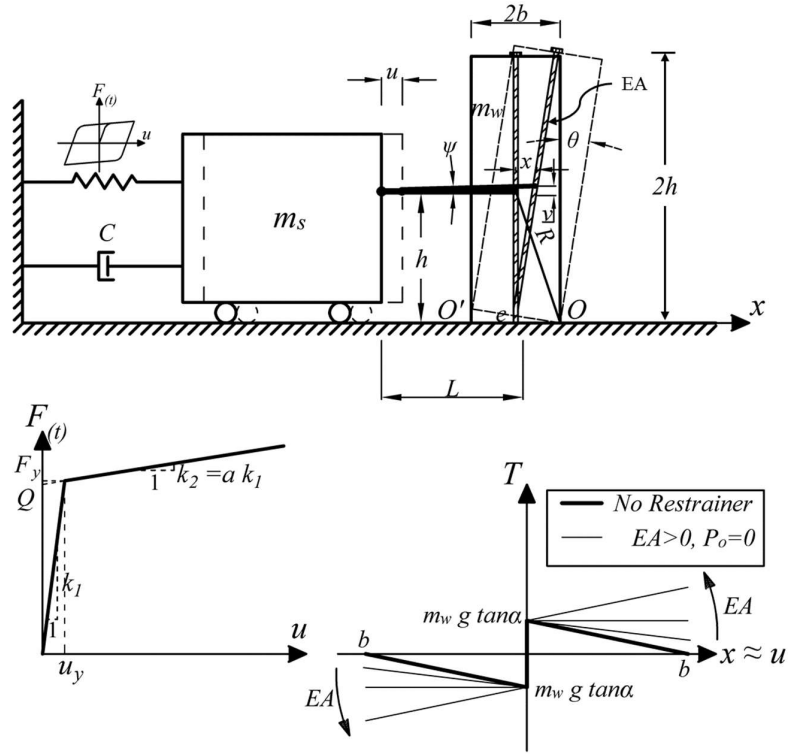

**Figure 1.** Yielding single-degree-of-freedom oscillator coupled with a vertically restrained stepping rocking wall.

Following these studies, Holden et al. (2003) presented experimental studies on the cyclic loading of a precast, partially prestressed system that incorporated post-tensioned unbonded tendons; while Ajrab et al. (2004) presented a performance-based design methodology for the frame-building-rocking-wall system with various prestressed tendon configurations and energy dissipation devices. In their proposed methodology Ajrab et al. (2004) adopt an "equivalent-static" lateral force procedure, and the study concludes that the proposed performance-based, capacity-demand method predicts larger displacements than those obtained from time-history analysis.

In the aforementioned studies, central postensioned steel tendons inside the rocking wall or bridge-pier are provided to increase the lateral resistance of the entire structure. The force-deformation curve of the vertically restrained solitary rocking wall reported in these studies has invariably a positive post-uplifting stiffness, indicating that the axial stiffness of the steel tendon is large enough to the extent that the post-uplift stiffness of the rocking wall is positive. By introducing such a stiff tendon that reverses the negative stiffness of the solitary rocking wall, one creates a strong system; nevertheless, at present it is not well understood to what extent a stiff vertical tendon that offers a positive lateral stiffness enhances the seismic stability of the overall structure or it merely contributes to accentuate the crushing of the pivoting points of the rocking wall due to the increased vertical load. Part of the motivation of this study is to build upon the previously referenced work and examine the role of vertical restrainers in the planar seismic response of moment-resisting frames coupled with rocking walls.

**DYNAMICS OF A YIELDING OSCILLATOR COUPLED WITH A VERTICALLY RESTRAINED STEPPING ROCKING WALL**

With reference to Figure (1), this study examines the dynamic response of a yielding single-degree-of-freedom (SDOF) structure, with mass, $m_s$, pre-yielding stiffness, $k_1$ post yielding stiffness, $k_2$, and strength, $Q$, that is coupled with a free-standing stepping rocking wall of size, $R = \sqrt{b^2 + h^2}$, slenderness, $\tan \alpha = b/h$, mass, $m_w$ and moment of inertia about the pivoting (stepping) points $O$ and $O'$, $I = 4/3\, m_w R^2$, that is vertically restrained with an elastic tendon with axial stiffness $EA$ which can be prestressed with a prestressing force $P_o$.

In the interest of simplicity, it is assumed that the arm with length, $L$, that couples the motion is articulated at the center of mass of the rocking wall at a height, $h$, from its foundation as shown in Figure 1.

During rocking motion, the center of mass of the rocking wall uplifts by $v$; therefore, the initially horizontal coupling arm rotates by an angle $\psi$. Accordingly, the horizontal translation of the center of mass of the rotating wall, $x$, is related to the horizontal displacement of the SDOF oscillator, $u$, via the expression, $\cos\psi = 1 - (u-x)/L$; whereas, $\sin\psi = v/L$. In this paper, the coupling arm is assumed to be long enough so that $v^2/L^2$ is much smaller that unity ($v^2/L^2 \ll 1$); and in this case $u=x$. A recent study by Makris and Aghagholizadeh (2017) on the response of an elastic oscillator coupled with a rocking wall showed that the effect due to a shorter coupling arm is negligible.

The system under consideration is a single-degree-of-freedom system where the lateral translation of the mass, $u$ is related to the rotation of the stepping rocking wall $\theta$ via the expression:

$$u = \pm R[\sin\alpha - \sin(\alpha \mp \theta)] \tag{1}$$
$$\dot{u} = R\dot{\theta}\cos(\alpha \mp \theta) \tag{2}$$
$$\ddot{u} = R[\ddot{\theta}\cos(\alpha \mp \theta) \pm \dot{\theta}^2\sin(\alpha \mp \theta)] \tag{3}$$

In equations (1) to (3), wherever there is a double sign (e.g. $\pm$) the top sign is for $\theta>0$ and the bottom sign is for $\theta<0$. Dynamic equilibrium of the mass $m_s$ gives:

$$m_s(\ddot{u} + \ddot{u}_g) = -F_s - c\dot{u} + T \tag{4}$$

where $F_s$ is the force the develops in the nonlinear spring and is described by the Bouc-Wen model (Wen 1976, Bouc 1967, Baber et al. 1981)

$$F_s(t) = ak_1 u(t) + (1-a)k_1 u_y z(t) \tag{5}$$

in which, $a = k_2/k_1$ is the post-to-pre yielding stiffness ratio and $-1 \leq z(t) \leq 1$ is a dimensionless internal variable described by:

$$\dot{z}(t) = \frac{1}{u_y}[\dot{u}(t) - \beta\dot{u}(t)|z(t)|^n - \gamma|\dot{u}|z(t)|z(t)|^{n-1}] \tag{6}$$

In equation (6), constants $\beta$, $\gamma$ and $n$ are model parameters. Furthermore, in equation (4), $T$ is the axial force (positive = tensile) that develops in the coupling arm.

During rocking motion of the vertically restrained wall, the tendon is elongated by (Vassiliou and Makris 2015)

$$e = \sqrt{2}R\sin\alpha\sqrt{1-\cos\theta} \tag{7}$$

In addition to the elongation, $e$, given by equation (7), the analysis accounts for an initial elongation

$$e_o = \frac{P_o}{EA/2h} \tag{8}$$

due to a possible initial postensioning force, $P_o$.
Accordingly, during rocking motion, the restoring moment on the rocking wall from the tendon alone is (Vassiliou and Makris 2015)

$$M_r = -R\sin\alpha\sin\theta\left[\frac{1}{2}EA\tan\alpha + \frac{P_o}{\sqrt{2}\sqrt{1-\cos\theta}}\right] \tag{9}$$

With reference to Figure 1 (bottom), as the elasticity of the tendon increases it offsets the negative stiffness originating from rocking. The value of the axial stiffness of the tendon that is needed to introduce positive stiffness is (Vassiliou and Makris 2015)

$$\frac{EA}{m_w g} = 2\frac{1}{\tan^2 \alpha} \tag{10}$$

For instance, for a slenderness value, $\tan \alpha = 1/6$, a rigid-plastic behavior is reached when $EA/(m_w g) = 72$.

*Case 1: θ>0*

For positive rotations ($\theta>0$), dynamic equilibrium of the vertically restrained stepping rocking wall with mass $m_w$ shown in Figure 1, gives:

$$I\ddot{\theta} = -TR\cos(\alpha - \theta) - m_w gR\sin(\alpha - \theta) - m_w \ddot{u}_g R\cos(\alpha - \theta)$$
$$-R\sin\alpha\sin\theta \left[\frac{1}{2}EA\tan\alpha + \frac{P_o}{\sqrt{2}\sqrt{1-\cos\theta}}\right] \tag{11}$$

where $P_o$ is the initial post-tensioning force and $EA$ is the axial stiffness of the elastic tendon. The axial force $T$ appearing in equation (11) is replaced with the help of equations (4) and (5), whereas for a rectangular stepping wall, $I = 4/3\, m_w R^2$. Accordingly, equation (11) assumes the form:

$$\frac{4}{3}m_w R^2 \ddot{\theta} + [m_s(\ddot{u} + \ddot{u}_g) + ak_1 u(t) + (1-a)k_1 u_y z(t) + c\dot{u}]R\cos(\alpha - \theta)$$
$$= -m_w R[\ddot{u}_g \cos(\alpha - \theta) + g\sin(\alpha - \theta)] - R\sin\alpha\sin\theta(\frac{1}{2}EA\tan\alpha + \frac{P_o}{\sqrt{2}\sqrt{1-\cos\theta}}) \tag{12}$$

Upon dividing with $m_w R$ equation (12) gives:

$$\frac{4}{3}R\ddot{\theta} + [\sigma(\ddot{u} + \ddot{u}_g) + a\frac{k_1}{m_w}u(t) + (1-a)\frac{k_1}{m_w}u_y z(t) + \frac{c}{m_w}\dot{u}]\cos(\alpha - \theta)$$
$$= -\ddot{u}_g \cos(\alpha - \theta) - g\sin(\alpha - \theta) - \sin\alpha\sin\theta(\frac{1}{2}\frac{EA}{m_w}\tan\alpha + \frac{P_o}{m_w\sqrt{2}\sqrt{1-\cos\theta}}) \tag{13}$$

in which $\sigma=m_s/m_w$ is the mass ratio parameter.
Substitution of the expressions of the relative displacement, velocity and acceleration given by equations (1) to (3) for positive rotations, and after dividing with $R$, equation (13) is expressed only in terms of the variable, $\theta(t)$.

$$(\frac{4}{3} + \sigma\cos^2(\alpha - \theta))\ddot{\theta} + \sigma\cos(\alpha - \theta)[a\omega_1^2(\sin\alpha - \sin(\alpha - \theta)) + 2\xi\omega_1 \dot{\theta}\cos(\alpha - \theta)$$
$$+\dot{\theta}^2 \sin(\alpha - \theta) + (1-a)\omega_1^2 \frac{u_y}{R} z(t)]$$
$$= -\frac{g}{R}[(\sigma + 1)\frac{\ddot{u}_g}{g}\cos(\alpha - \theta) + \sin(\alpha - \theta) + \sin\alpha\sin\theta(\frac{1}{2}\frac{EA}{m_w g}\tan\alpha + \frac{P_o}{m_w g}\frac{1}{\sqrt{2}\sqrt{1-\cos\theta}})] \tag{14}$$

where $\omega_1 = \sqrt{k_1/m_s}$ = the pre-yielding undamped frequency and $\xi = \frac{c}{2m_s\omega_1}$ = the pre-yielding viscous damping ratio of the SDOF oscillator. Equation (14) is the equation of motion for positive rotations of the coupled system shown in Figure 1.

*Case 2: θ<0*

For negative rotations one can follow the same reasoning and the equation of the coupled system shown in Figure 1 is:

$$(\frac{4}{3} + \sigma\cos^2(\alpha + \theta))\ddot{\theta} - \sigma\cos(\alpha + \theta)[a\omega_1^2(\sin\alpha - \sin(\alpha + \theta)) - 2\xi\omega_1 \dot{\theta}\cos(\alpha + \theta)$$
$$+\dot{\theta}^2 \sin(\alpha + \theta) - (1-a)\omega_1^2 \frac{u_y}{R} z(t)]$$
$$= \frac{g}{R}[-(\sigma + 1)\frac{\ddot{u}_g}{g}\cos(\alpha + \theta) + \sin(\alpha + \theta) - \sin\alpha\sin\theta(\frac{1}{2}\frac{EA}{m_w g}\tan\alpha + \frac{P_o}{m_w g}\frac{1}{\sqrt{2}\sqrt{1-\cos\theta}})] \tag{15}$$

When parameters $EA/m_w g=P_o/m_w g=0$, equations (14) and (15) collapse to the equations of motion presented in (Makris and Aghagholizadeh 2017, Aghagholizadeh and Makris 2018) for a yielding SDOF oscillator coupled with a rocking wall with no vertical restrainer. The terms multiplied with the parameter $\sigma=m_s/m_w$ are associated with the dynamics of the yielding SDOF oscillator; whereas, the remaining terms are associated with the dynamics of the rocking wall. When the SDOF oscillator is absent ($\sigma=\omega_1=\xi=0$), equations (14) and (15) reduce to the equations of motion of the solitary restrained rocking wall (Vassiliou and Makris 2015) since the frequency parameter p for rectangular walls is $p = \sqrt{3g/4R}$ (Makris 2014a, b, Makris and Kampas 2016). Equations (14) and (15) reveal that the effect of the tendon ($EA$ and $P_o$) is different than the effect of a heavier wall (lower σ). These differences are illustrated in the response spectra presented later in the paper. The energy loss during impact is a function of the wall-foundation interface. Accordingly, the coefficient of restitution, $e = \dot{\theta}_2/\dot{\theta}_1 < 1$, is introduced as a parameter of the problem. In this study the coefficient of restitution assumes the value of *e= 0.9*. The uplift acceleration of the system is:

$$\ddot{u}_g = \frac{1}{\sigma + 1} g \tan\alpha (1 + \frac{P_o}{m_w g}) \tag{16}$$

(Makris and Aghagholizadeh 2017, Aghagholizadeh and Makris 2018).

## PARAMETERS OF THE PROBLEM

The Bouc-Wen model described by equations (5) and (6) is a phenomenological model of hysteresis originally proposed by Bouc (1967) and subsequently generalized by Wen (1976, 1975) and Baber and Wen (1981). It is a versatile model that can capture various details of the nonlinear force-displacement loop. Subsequent studies on the modeling of yielding systems by Constantinou and Adnane (1987) concluded that when certain constraints are imposed on the parameters β and γ (β+γ=1), the model reduces to a viscoplastic element with well-defined physical characteristics. The Bouc-Wen model essentially builds on the bilinear idealization shown in the bottom-left of Figure 1.

While the five parameters appear in the bilinear idealization. ($k_1$= pre-yielding stiffness, $k_2$= post-yielding stiffness, $u_y$= yield displacement, $Q$ = strength and $F_y$= yielding force), only three parameters are needed to fully describe the bilinear behavior (see for instance Makris and Kampas 2013). In this work, the authors select the pre-yielding stiffness $k_1 = m_s \omega_1^2$, the post-yielding stiffness $k_2=ak_1$ and the strength of the structure $Q$. With reference to Figure 1 (bottom-left), $F_y = k_1 u_y = Q + k_2 u_y$. Accordingly, $u_y = Q/(k_1 - k_2)$ and $F_y = k_1 Q/(k_1 - k_2)$. The parameters *β, γ* and *n* appearing in equation (6) are established from past studies on the parameter identification of yielding concrete structures and assume the values: *β=0.95, γ=0.05* and *n=2* (Kunnath et al. 1997, Goda et al. 2009). With the parameters *β=0.95, γ=0.05* and *n=2* being established, the peak inelastic displacement, $u_{max}$ of the SDOF idealization shown in Figure 1 is a function of the following parameters:

$$u_{max} = f(\omega_1, \frac{Q}{m_s}, a, \xi, p, \tan\alpha, \sigma, g, EA, P_o, parameters\ of\ excitation) \tag{17}$$

In this study, it is assumed that upon yielding, the structure maintains a mild, positive, post-yielding stiffness $= k_2 = 0.05 k_1$, therefore $a = 0.05$ (Kunnath et al. 1997, Goda et al. 2009). Furthermore, it is assumed that the pre-yielding damping ratio, $\xi = c/(2m_s \omega_1) = 0.03$ and the paper focuses on rocking walls with slenderness, $\tan \alpha = 1/6$.

## VALIDATION OF THE SDOF – IDEALIZATION

In this section the dependability of the single-degree-of-freedom idealization shown in Figure 1 is examined against the results obtained with the open-source code OpenSees (2000) when analyzing the nine-story moment resisting steel structure designed for the SAC Phase II Project (2000). This structure that is well-known to the literature (Gupta and Krawinkler 1999, Chopra and Goel 2002) was designed to meet the seismic code (pre-Northridge Earthquake) and represents typical medium-rise buildings designed for the greater area of Los Angeles, California.

This moment-resisting, steel building is 40.82 m tall with 9-stories above ground level and a basement. The bays are 9.15 m wide, with five bays in north-south (N-S) and east-west (E-W) directions. Floor-to-floor height of each story is 3.96 m, except for the basement and first floor which are 3.65 m and 5.49 m respectively as shown in Figure 2. Columns splices are on the 1st, 3rd, 5th and 7th floors and located 1.83 m above the beam-column joint. The column bases are modeled as pinned connections and it is assumed that the surrounding soil and concrete foundation walls are restraining the structure in horizontal direction at the ground level. The columns are 345 MPa wide-flange steel sections and the floor beams are composed of 248 MPa wide-flanges steel sections. All beam column connections of the frames are rigid except for the corner columns which are pinned in order to avoid bi-axial bending of the members. In this study, the exterior frame in N-S direction is chosen for the 2-D validation of our planar analysis.

The nonlinear response of the nine-story MDOF structure is computed with the nonlinear built-in model "Steel01" in OpenSees. It is a bilinear model at the stress-strain level with a backbone curve that is similar to the force-displacement curve shown in Figure 1. Accordingly, we have used an elastic modulus of $E=210$ GPa, a strain hardening ratio (post-yield to elastic, pre-yield modulus ratio), a=0.03 and a yield strength, $\sigma_y=248$ MPa for beams and $\sigma_y=345$ MPa for columns.

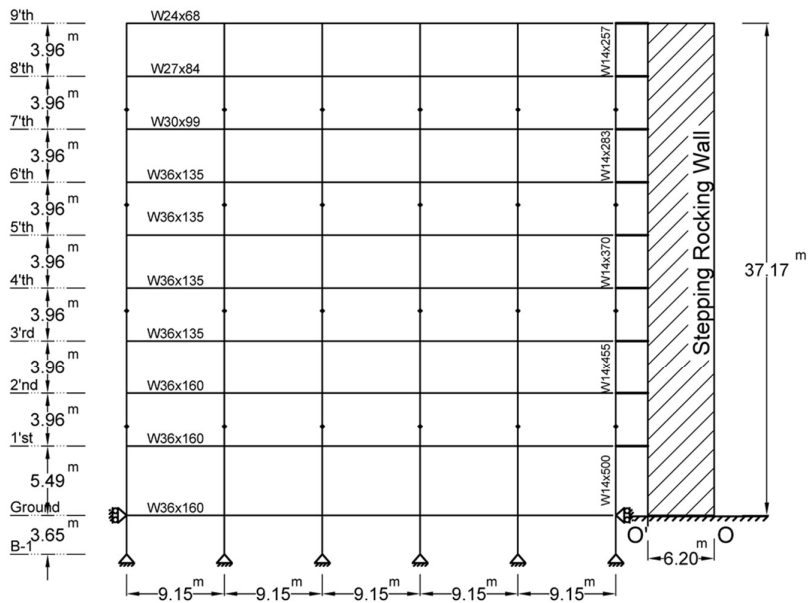

**Figure 2.** Top: Nine-story moment-resisting steel frame designed for the SAC Phase II Project coupled with a stepping rocking wall. Bottom: Geometric and physical characteristics pertinent to the 9-story SAC building. The indicated seismic mass is the entire mass of each floor of the SAC building. Given that the 9-story SAC building has two parallel moment-resisting frames in the direction of shaking, the mass $m_s = 4.5 \times 10^6 kg$ is half the seismic mass indicated in Figure 2.

Figure 3 (top) plots the computed push-over curve (base shear vs roof displacement) of the 9-story moment resisting steel building without rocking wall, which is essentially identical with the push-over curve presented in past investigations (Gupta and Krawinkler 1999, Chopra and Goel 2002). From the push-over curve shown in Figure 3 (a), the strength of the nine-story frame is estimated to be approximately $Q = V_y = 7616 kN$ and is reached at a yield displacement $u_y = 36.2 cm$. These values are used for the strength and yield displacement of the SDOF bilinear model shown in Figure 1. The resulting pre-yielding period of the building is $T_1 = 2.27\ sec$, while its normalized strength is $Q/m_s = 0.17g$. The force distribution used to produce the push-over curve of the nine-story moment resisting yielding frame is the first-mode force-distribution presented in Figure 8 of reference (Chopra and Goel 2002).

The remaining four subplots in Figure 3 plot the base-shear versus the mid-height displacement of the 9-story building without rocking wall together with the corresponding force-displacement loops computed with Matlab of the SDOF inelastic model shown in Figure 1 when excited with the 1994 Newhall/360, Northridge (b), 1992 Erzincan NS, Turkey (c), the 1995 Takarazuka/000, Kobe (d) and the 1971 Pacoima Dam/164, Imperial Valley (e) ground motions. The earthquake excitation is induced at the fixed nodes of the structural model shown in Figure 2—that is the supports of the columns at the basement level and at the base of the stepping wall at the ground level. All four subplots show that the inelastic force-displacement loops of the SDOF model shown in. Figure 1 follow with fidelity the inelastic back-bone curve of the 9-story SAC building that is computed with OpenSees (2000).

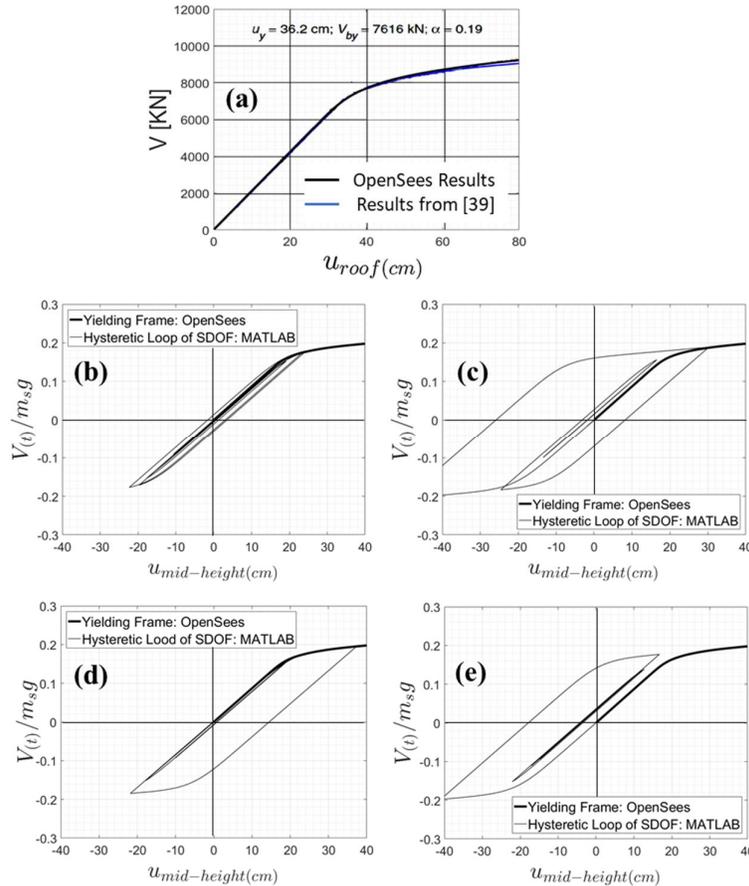

**Figure 3.** (a): Comparison of the computed push-over curve (base-shear vs roof displacement) of the 9-story moment-resisting steel building with the results reported by (Chopra and Goel 2002). Base-shear versus displacement at mid-height computed with OpenSees of the 9-story steel building without rocking wall together with the corresponding force-displacement loops computed with MATLAB of the SDOF inelastic model shown in Figure 1 when excited with the 1994 Newhall/360, Northridge (b), the 1992 Erzincan NS, Turkey (c), the 1995 Takarazuka/000, Kobe (d) and the 1971 Pacoima Dam/164, Imperial Valley (e) ground motions.

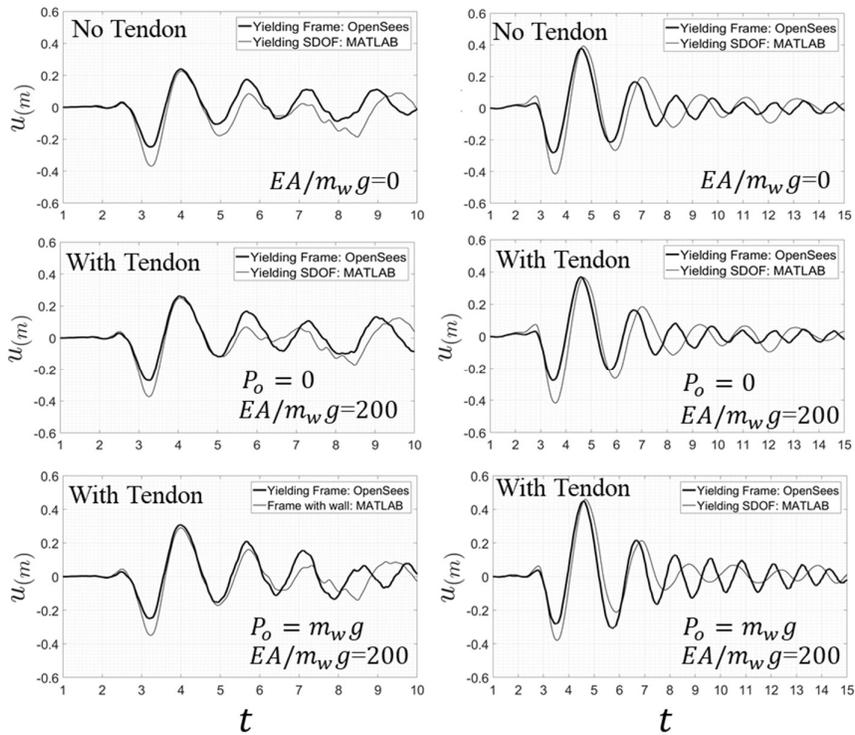

**Figure 4.** Comparison of the displacement time histories at mid-height of the 9-story steel building shown in Figure 2, computed with OpenSees with the displacement time-histories of the SDOF idealization shown in Figure 1, when excited with the 1971 Pacoima Dam/164, San Fernando, California (left) and the 1992 Erzincan NS, Turkey (right) ground motions.

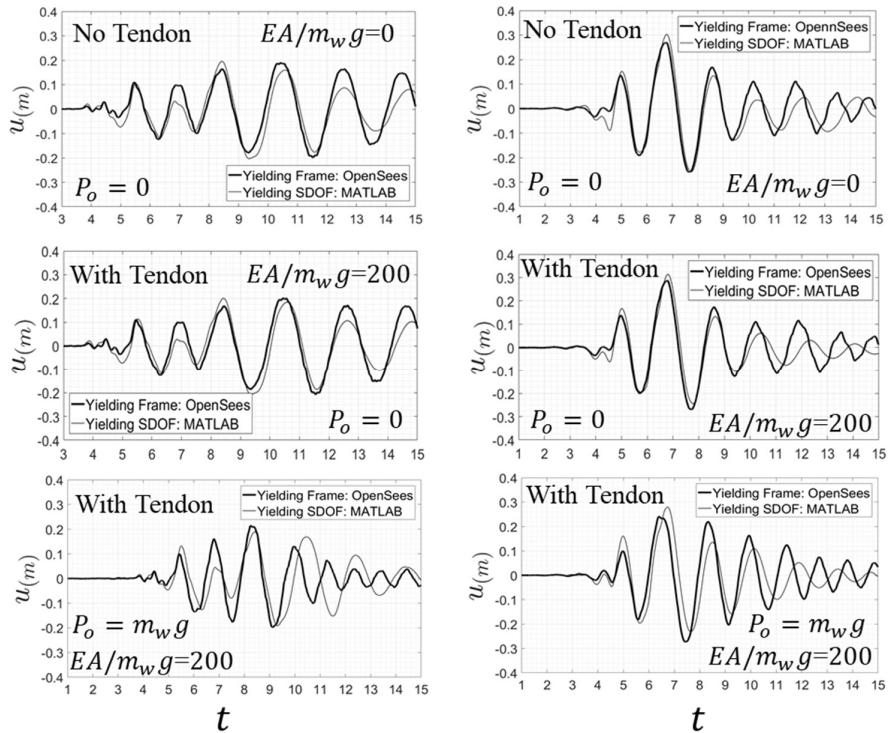

**Figure 5.** Comparison of the displacement time histories at mid-height of the 9-story steel building shown in Figure 2, computed with OpenSees with the displacement time-histories of the SDOF idealization shown in Figure 1, when excited with the 1994 Newhall/360, Northridge, California (left) and the 1995 Takarazuka/000, Kobe, Japan (right) ground motions.

When analyzing with OpenSees the 9-story SAC building coupled with the stepping rocking wall as shown in Figure 2, the properties of the rocking interface are approximated with a rigid-elastic rotational spring together with a rotational viscous dashpot (McKenna et al. 2000) to approximate the energy loss during impact as the rocking wall alternate pivot-points. For a free-standing stepping rocking wall with size $R$, slenderness, $\alpha$, and mass $m_w$, dimensional analysis yields that the expression of the equivalent rotational dashpot is (Vassiliou et al. 20014)

$$c = \lambda \alpha^2 m_w g^{0.5} R^{1.5} \qquad (31)$$

where, $\lambda = 250$ is a parameter that is calibrated from best fit of the result.

Figure 4 compares response histories computed with OpenSees at mid-height of the 9-story SAC steel building with the solutions obtained with MATLAB for the SDOF idealization shown in Figure1. The top plots are when the rocking wall is not restrained (No tendon), the center plots are when the rocking wall is restrained with a stiff tendon with $EA/m_w g = 200$ without being prestressed ($P_o = 0$); while, the bottom plots are when the tendon with $EA/m_w g = 200$ is prestressed with $P_o = m_w g$. The left plots are when the structure is subjected to the Pacoima Dam/164 ground motion recorded during the 1971 San Fernando, California earthquake whereas the right plots are when the structure is subjected to the Erzincan NS ground motion recorded during the 1992 Erzincan, Turkey earthquake. The comparison of the OpenSees and Matlab solutions are in good agreement—in particular for the peak-response values and supports the use of the SDOF idealization introduced in Figure 1.

Equally good comparisons are plotted in Figure 5 when the inelastic structure coupled with the rocking wall is subjected to the Newhall/360 ground motion recorded during the 1994 Northridge, California earthquake (left plots) and the Takarazuka/000 ground motion recorded during the 1995 Kobe, Japan earthquake.

**EARTHQUAKE SPECTRA OF A YIELDING OSCILLATOR COUPLED WITH A ROCKING WALL**

Following the verification of the single-degree of freedom idealization shown in Figure 1 by comparing its response with that of the 9-story steel SAC building computed with OpenSees, the equations of motion (14) and (15) are used to generate inelastic response spectra. Figure 6 plots displacement spectra of a yielding SDOF oscillator coupled with a vertically prestressed, stepping rocking wall when excited by the Newhall/360 ground motion recorded during the 1994 Northridge, California earthquake.

The left plots are for a structure with a yielding strength $Q/m_s = 0.15g$; whereas, the right plots are for a weaker structure, $Q/m_s = 0.08g$. The first and most important observation is that the effect of vertical tendons even when they are stiff ($\frac{EA}{m_w g} = 200$) and highly prestressed ($p_o = m_w g$) is marginal. In contrast, the weight of the rocking wall has more noticeable effects with the heavier wall ($\sigma = 5$) being more effective in some regions of the spectra. The same conclusions are drawn from the inelastic spectra presented in Figure 7 and 8 where the inelastic structure-rocking wall system is excited by the Pacoima Dam/164 ground motion recorded during the 1971 San Fernando, California earthquake and the Erzincan NS ground motion recorded during the 1992 Erzincan, Turkey earthquake.

**CONCLUSIONS**

This paper investigates the dynamic response of a yielding SDOF oscillator coupled with a vertically restrained, stepping rocking wall. The full nonlinear equations of motion were derived, and the dependability of the one-degree-of-freedom idealization is validated against the nonlinear time-history response analysis of the 9-story SAC steel building. The equations of motion of the SDOF idealization show explicitly that the contribution of vertical tendons, even when they are stiff, is two orders of magnitude less than the inertia forces on the moment frame-rocking wall system. This paper offers a comprehensive parametric analysis which reaches the following conclusions.

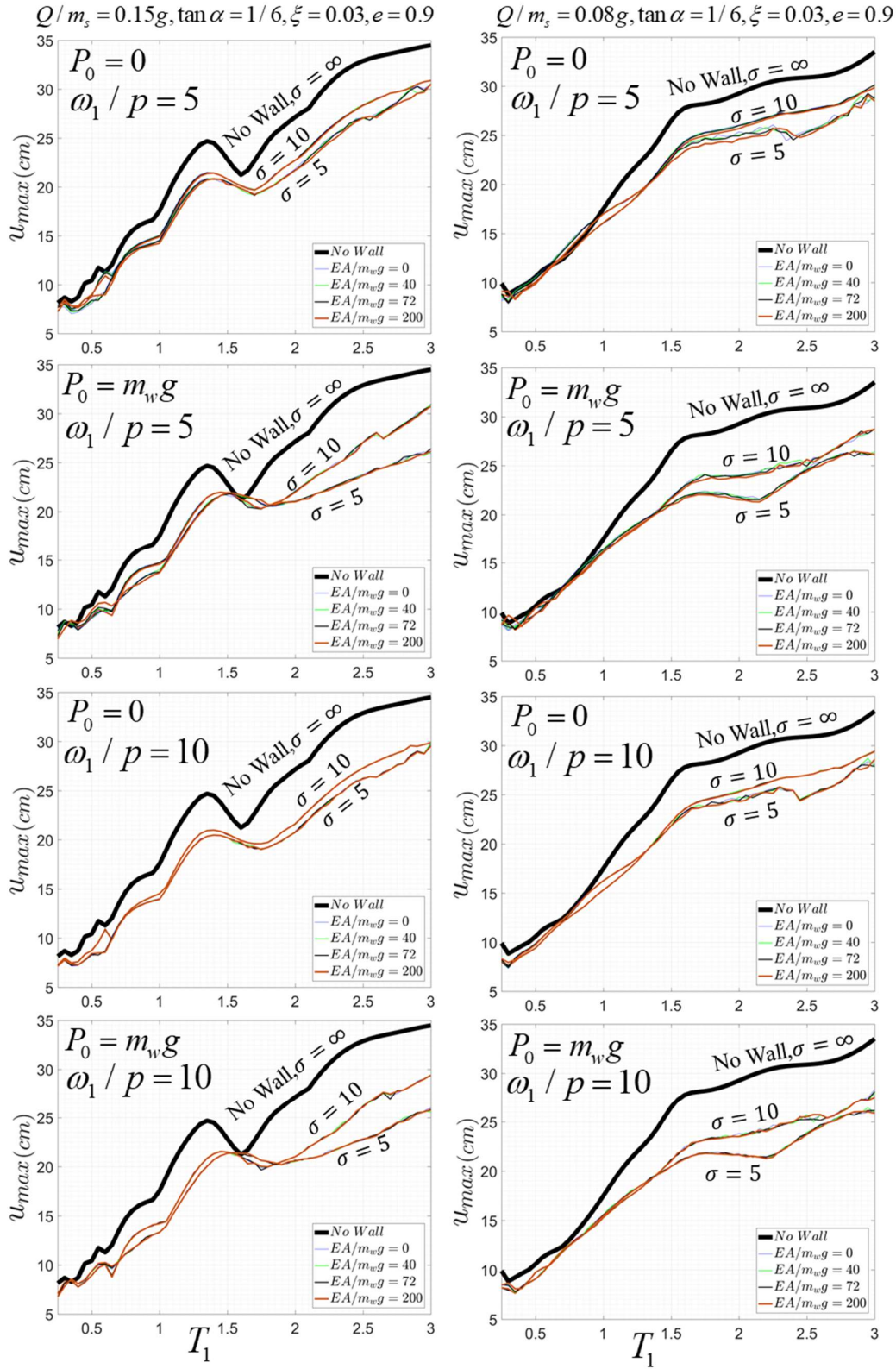

**Figure 6.** Displacement spectra of a yielding SDOF oscillator coupled with a vertically restrained stepping rocking wall with slenderness $\tan\alpha = 1/6$, for two valued of strength, $Q/m_s = 0.15g$ (left column) and $Q/m_s = 0.08g$ (right column) with mass ratios, $\sigma = 5, 10$ and $\infty$ (no wall); several values of tendon stiffness ($EA/m_w g$=0, 40, 72 and 200) with ($P_o = m_w g$) and without ($P_o = 0$) pre-tensioning when subjected to the Newhall/360 ground motion recorded during the 1994, Northridge California earthquake.

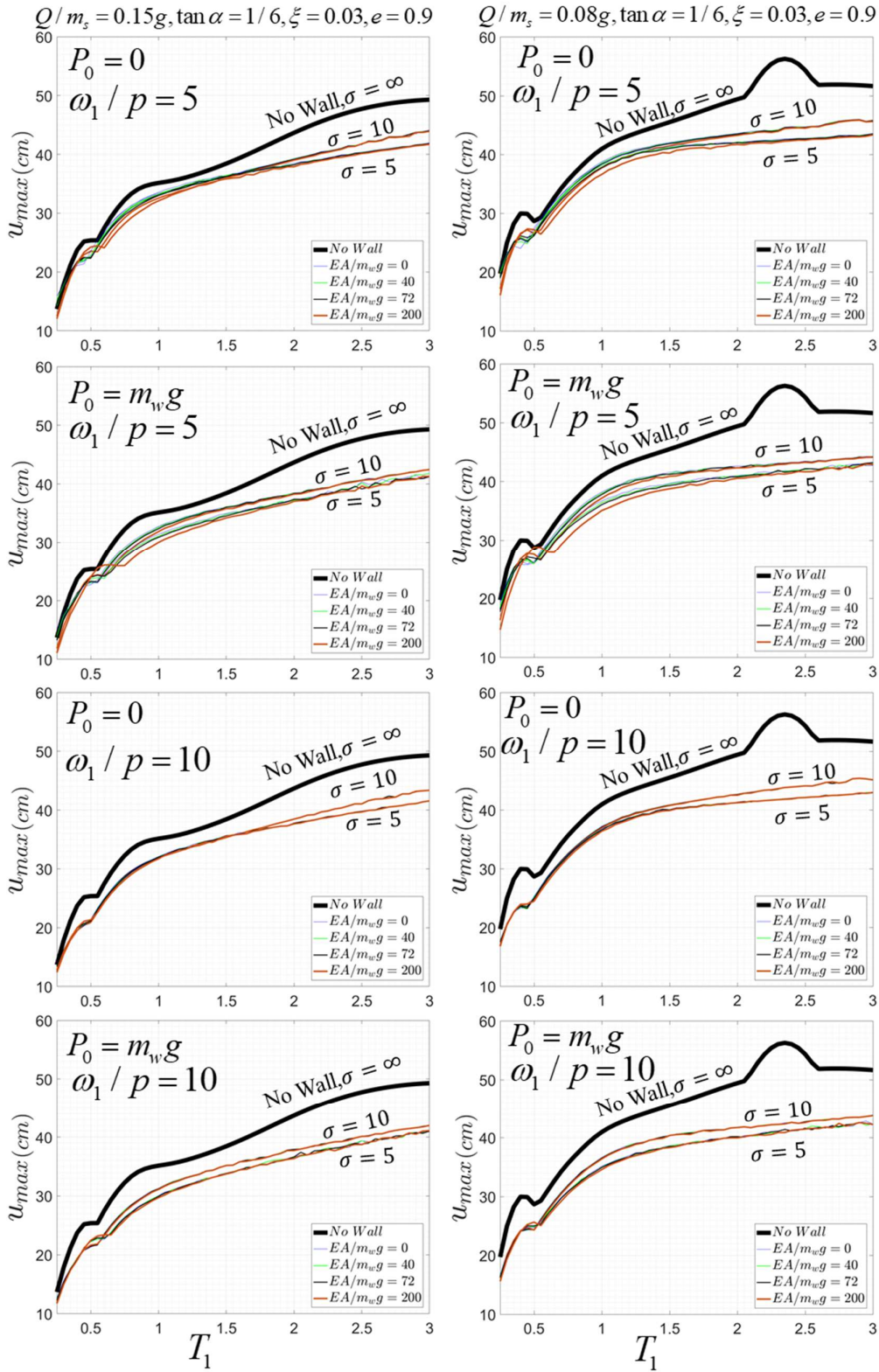

**FIGURE 7.** Displacement spectra of a yielding SDOF oscillator coupled with a vertically restrained stepping rocking wall with slenderness $\tan\alpha = 1/6$, for two valued of strength, $Q/m_s = 0.15g$ (left column) and $Q/m_s = 0.08g$ (right column) with mass ratios, $\sigma = 5, 10$ and $\infty$ (no wall); several values of tendon stiffness ($EA/m_w g$=0, 40, 72 and 200) with ($P_o = m_w g$) and without ($P_o = 0$) pre-tensioning when subjected to the Pacoima Dam/164 ground motion recorded during the 1971 San Fernando, California earthquake.

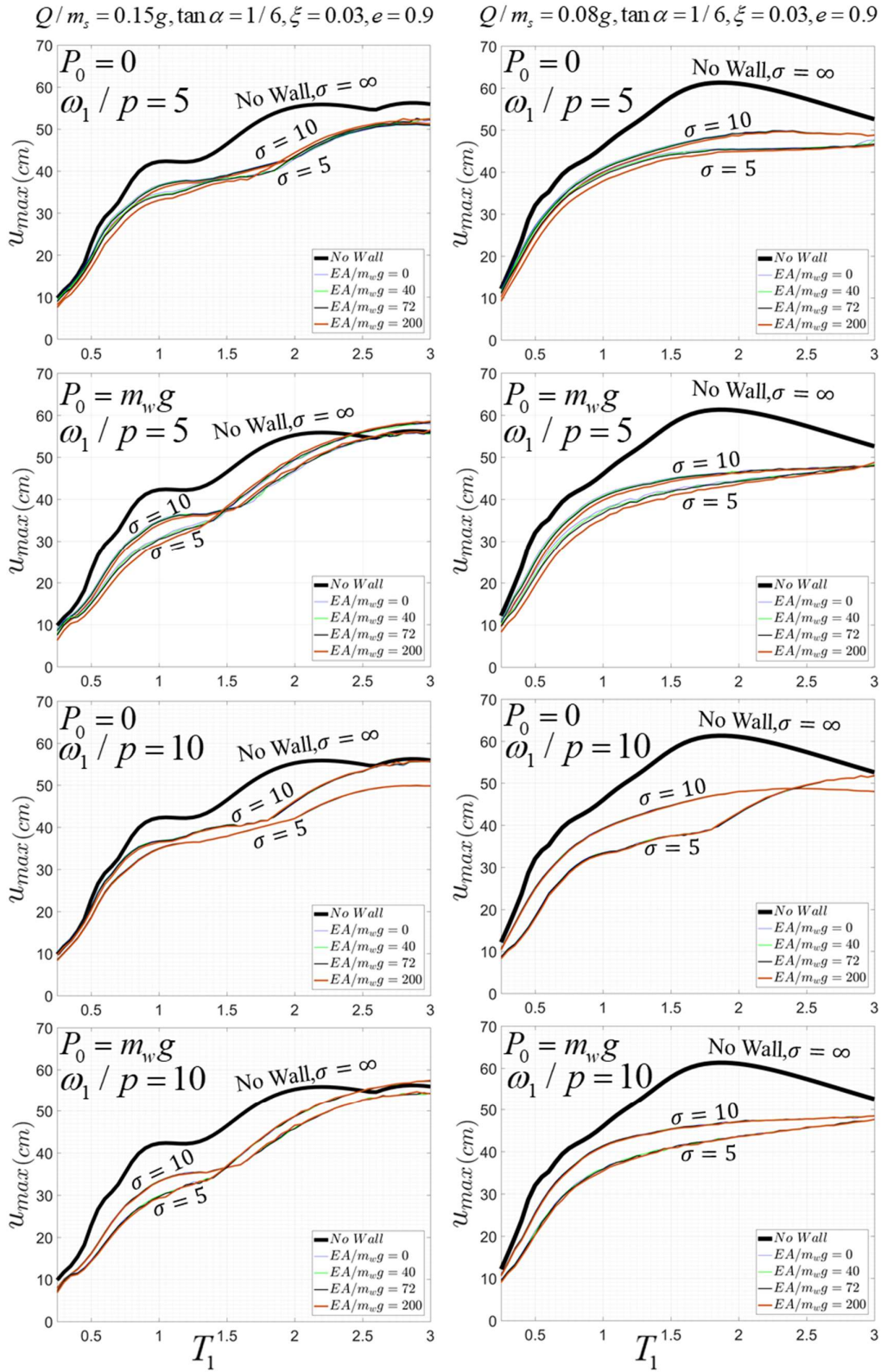

**FIGURE 8.** Displacement spectra of a yielding SDOF oscillator coupled with a vertically restrained stepping rocking wall with slenderness $\tan\alpha = 1/6$, for two valued of strength, $Q/m_s = 0.15g$ (left column) and $Q/m_s = 0.08g$ (right column) with mass ratios, $\sigma = 5, 10$ and $\infty$ (no wall); several values of tendon stiffness ($EA/m_w g$=0, 40, 72 and 200) with ($P_o = m_w g$) and without ($P_o = 0$) pre-tensioning when subjected to the Erzincan NS ground motion recorded during the 1992 Erzincan, Turkey earthquake.

The participation of the stepping rocking wall suppresses peak inelastic displacements with the heavier wall being in most cases more effective. In contrast, the effect of the vertical tendons even when they are stiff ($\frac{EA}{m_w g} = 200$) and highly prestressed ($P_o = m_w g$) is marginal. Given than the vertical tendons increase the vertical reactions at the pivoting corners by more than 50%, the paper concludes that for medium- to high-rise buildings, vertical tendons in rocking walls are not recommended.

The SDOF idealization presented in this paper compares satisfactory with finite-element analysis of a 9-story steel SAC building coupled with a stepping rocking wall; therefore, the SDOF idealization can be used with confidence for preliminary analysis and design.